\documentclass[12pt]{article} 
\usepackage{epsfig}
\usepackage{amssymb}

\renewcommand{\thefootnote}{\fnsymbol{footnote}}

\newcommand{\Kperp}{K^*_\perp}
\newcommand{\Kpar}{K^*_\parallel}
\newcommand{\qz}{(q\cdot z)}
\newcommand{\ms}{\overline{m}_s(\mu)}
\newcommand{\ub}{\bar u}
\newcommand{\quark}{\langle \bar q q\rangle}
\newcommand{\mixed}{\langle \bar q \sigma gG q\rangle}
\newcommand{\squark}{\langle \bar s s\rangle}
\newcommand{\smixed}{\langle \bar s \sigma gG s\rangle}
\newcommand{\gluon}{\left\langle \frac{\alpha_s}{\pi}\,G^2\right\rangle}

\begin{document}

\begin{titlepage}
\begin{flushright}
\begin{tabular}{l}
IPPP/03/44 \\
DCPT/03/88\\
July 2003
\end{tabular}
\end{flushright}
\vskip1.5cm
\begin{center}
   {\Large \bf SU(3) Breaking in K and K$^*$
    Distribution Amplitudes}
    \vskip1.3cm {\sc
Patricia Ball\footnote{Patricia.Ball@durham.ac.uk} and 
    M. Boglione\footnote{Elena.Boglione@durham.ac.uk}}
  \vskip0.2cm
	IPPP, Department of Physics, 
University of Durham, Durham DH1 3LE, UK\\ 
  \vskip2cm


\vskip5cm

{\large\bf Abstract:\\[10pt]} \parbox[t]{\textwidth}{ 
We calculate the decay couplings and first two Gegenbauer moments of
the leading twist light-cone distribution amplitudes of K and K$^*$
from QCD sum rules, including NLO perturbative effects.
}

\vskip5cm

{\em  Submitted to Physical Review D}
\vfill
\end{center}
\end{titlepage}

\renewcommand{\theequation}{\arabic{section}.\arabic{equation}}
\renewcommand{\thefootnote}{\arabic{footnote}}
\setcounter{footnote}{0}

\section{Introduction and Motivation}
\setcounter{equation}{0}

Hadronic light-cone distribution amplitudes (DAs) of 
leading twist play an essential
r\^{o}le in the QCD description of hard exclusive processes. They can
be obtained from the hadron's Bethe-Salpeter equation by integration
over the transverse momentum distribution, but keeping track of the
longitudinal momentum fraction $u$:
$$
\phi(u) \sim \int_{k^2_\perp <\mu^2}d^2k_\perp\phi(u,k_\perp).
$$
DAs enter the amplitudes of processes to which collinear factorisation
theorems apply and which notably include ``classical'' applications like
the EM form factor of the pion or the $\gamma\gamma^*$-$\pi$
transition form factor, which were first discussed in the seminal
papers by Brodsky, Lepage and others \cite{exclusive}. More recently,
collinear factorisation has been shown to apply, to leading order in
an expansion in $1/m_b$, also to a large class
of nonleptonic B decays \cite{BBNS}, which has opened a new and
exciting area of
applications of meson DAs: nonleptonic B decays, and in particular
CP asymmetries in these decays, are currently being studied at the B factories
BaBar and Belle and are expected to yield essential information about the
pattern of CP violation and potential sources of flavour violation
beyond the SM. 
DAs also enter crucially SCET, the soft-collinear
effective theory \cite{SCET}, 
which aims to provide a unified theoretical framework for the
factorisation of both hard-collinear effects, relevant for all hard
exclusive QCD processes, and soft effects, which occur in processes
involving heavy mesons.

For light mesons, on which we shall concentrate in this paper, 
leading twist DAs can be interpreted as probability amplitudes of finding
the meson in a state with a minimum number of Fock constituents and at
small transverse separation which provides an ultraviolet cut-off. The
dependence on the this cut-off $\mu$ is given by Brodsky-Lepage 
evolution equations and can be calculated in perturbative QCD, while
the DAs at a certain low scale provide the
necessary nonperturbative input for a rigorous QCD treatment of
exclusive reactions with large momentum transfer.

As exclusive QCD processes come with much smaller cross-sections or
branching ratios than inclusive processes, they are not as well
studied experimentally as their inclusive counterparts. The overall
normalisations of DAs are given by local hadronic matrix elements,
essentially decay constants, which are partly accessible
experimentally,
partly have to be calculated from theory. Information on
the shape of DAs comes mostly from theory and has
been the subject of numerous studies within various nonperturbative
approaches. There are a few, mostly exploratory, studies of the second 
momentum of the pion DA in lattice QCD, Ref.\ \cite{latticeDAs}, but
the results are not yet refined enough
 to be relevant for phenomenology. Most of the existing 
information on light meson DAs comes from QCD sum
rules, whose application to the study of moments of DAs has been
pioneered by Chernyak and Zhitnitsky. Their review
on the ``Asymptotic 
Behavior Of Exclusive Processes In QCD'', Ref.~\cite{CZreport}, 
is still a very useful source
of information, despite having been written 20 years ago.
Later analyses switched
from the calculation of moments to that of {\em Gegenbauer} moments
which, as we shall discuss in Sec.~2, are more appropriate for the
study of DAs. The leading-twist DA of the pion was analysed in 
Ref.~\cite{filyanov}, while those of the $\rho$ were studied in
Ref.~\cite{BB96}. The first paper to study SU(3) breaking effects
in moments was Ref.~\cite{Russians}, while the first two Gegenbauer moments 
of the $K^*$ and $\Phi$ meson
DAs were calculated in Ref.~\cite{BBKT}.

The aim of the present paper is to provide a careful analysis of
SU(3) breaking effects in leading-twist $K$ and $K^*$ DAs, using QCD
sum rules. The motivation for
this study is the need for a more accurate assessment of these
effects than presented in the papers quoted above. This need is driven
by the fact that DAs are crucial ingredients in the
analysis of B decays of type $B\to\pi K$ etc., 
which are presently being measured
at the B factories. In the future, accurate information on
SU(3) breaking effects in DAs will also be required for assessing the
reliability of phenomenological methods for relating, via U-spin
symmetry, amplitudes of $B_d$ decays to those of $B_s$ decays which
will be measured
at the Tevatron and the LHC. SU(3) breaking enters these processes in
essentially three different ways:
\begin{itemize}
\item via an asymmetry of the DAs under the exchange of quark and
  antiquark;
\item via SU(3) breaking in the symmetric parts of the DA, in
  particular the overall normalisation;
\item via form factors, e.g.\ $F^{B\to\pi}$ vs.\ $F^{B\to K}$.
\end{itemize}
Although in the present paper we will concentrate on the first two 
manifestations
of SU(3) breaking, our results can be used immediately for an update
of $B\to K,K^*$ form factors using QCD sum rules on the light-cone
(cf.~\cite{FFs} for the present state of the art).

Our paper is organised as follows: Sec.~2 contains the definition of
all relevant DAs as well as a discussion of their behaviour under a
change of renormalisation scale. In Sec.~3 we discuss QCD sum rules for
the DAs, obtain numerical results and
compare with the results of other studies. Section~4 presents a
summary and our conclusions. Issues of a more technical nature are
discussed in the appendices.

\section{General Framework}
\setcounter{equation}{0}
\subsection{Definitions}
  We define the light-cone DAs via matrix elements of
  quark-antiquark gauge-invariant nonlocal operators at light-like
  separations $z_\mu$ with $z^2=0$ \cite{CZreport}. For definiteness
  we consider the
  $K^{(*)-}$ meson distributions; the DAs of the neutral mesons
  containing an s quark just
  involve a trivial isospin factor $1/\sqrt{2}$ in the overall
  normalisation. For mesons containing an s antiquark, one has 
  $\phi_{(\bar s q)}(u) =  \phi_{(\bar q s)}(1-u)$. The complete set of 
  distributions to leading-twist accuracy involves three DAs (we use
  the notation $\hat{z} = z^\mu\gamma_\mu$ for arbitrary 4-vectors $z$):
  \begin{eqnarray}
    \langle 0 |\bar u(z)\hat{z}\gamma_5 [z,0]s(0)
  |K^{-}(q)\rangle &=& i  f_K \qz 
  \int_0^1 du\, e^{-i\ub\qz} \phi_K(u)\,,\nonumber\\
  \langle 0 |\bar u(z)\hat{z}[z,0] s(0)
  |K^{*-}(q,\lambda)\rangle &=& (e^{(\lambda)} z)
  f_K^\parallel m_{K^*}\int_0^1 du\, e^{-i\ub\qz}
  \phi_K^\parallel(u),\nonumber\\
  \langle 0 |\bar u(z)\sigma_{\mu\nu}[z,0]s(0)
  |K^{*-}(q,\lambda)\rangle & = &
  i(e^{(\lambda)}_\mu q_\nu -e^{(\lambda)}_\nu q_\mu)
  f_K^\perp(\mu) \int_0^1 du\, e^{-i\ub\qz} \phi_K^\perp(u),
  \label{eq:defDAs}
  \end{eqnarray}
  with the Wilson-line
  $$
  [z,0] = \mbox{Pexp}\,\left[ig\int_0^1 d\alpha\, z^\mu A_\mu(\alpha
    z)\right]
  $$
inserted between quark fields to render the matrix elements
gauge-invariant. 
  In the above definitions, $e^{(\lambda)}_\nu$ is the
   polarization vector of a vector meson with polarisation
   $\lambda$; there are two DAs for vector mesons, corresponding to
   longitudinal and transverse polarisation, respectively.
 The integration variable $u$ is the
meson momentum fraction carried by the quark, $\ub \equiv 1-u$ the
  momentum fraction carried by the antiquark.  The normalisation constants
  $f_K$ are defined by the local limit of
  Eqs.~(\ref{eq:defDAs}) and chosen in such a way that 
  \begin{equation}\label{eq:norm} \int_0^1 du\, \phi(u)=1\end{equation}
  for all the three distributions $\phi_K,
  \phi_K^\parallel , \phi_K^\perp$. 

  \subsection{Conformal Expansion and Renormalisation}

The conformal expansion of light-cone distribution
amplitudes is analogous to the partial wave expansion
of  wave functions in standard quantum mechanics.   
In conformal expansion, 
the invariance of massless QCD under conformal transformations
is the equivalent of rotational symmetry in quantum mechanics, where,
for spherically symmetric potentials, the partial wave 
decomposition serves to separate angular degrees of freedom 
from radial ones. All dependence on the angular 
coordinates is included in spherical harmonics which form an irreducible
representation of the group O(3), and the dependence on the single 
remaining radial coordinate is governed by a one-dimensional 
Schr\"{o}dinger equation. Similarly, the conformal expansion of distribution 
amplitudes in QCD aims to separate longitudinal degrees of freedom 
from  transverse ones. All dependence on the longitudinal momentum fractions 
is described by orthogonal polynomials 
that form an irreducible representation of the so-called collinear subgroup 
of the conformal group, SL(2,$\mathbb R$), describing M\"obius
transformations on the 
light-cone. The transverse-momentum dependence
(scale-dependence) is governed by simple renormalisation group equations:
the different partial waves,
labelled by different ``conformal spins'',
behave independently and do not mix with each other.
Since the conformal invariance of QCD is 
broken by quantum corrections, mixing between different terms of the 
conformal expansion is absent only to leading logarithmic accuracy.
Still, conformal spin is a good quantum number in hard processes,
up to small corrections of order $\alpha_{s}^{2}$. 
The application of conformal symmetry to the study of exclusive processes
to leading twist has become a vast field whose current status is
reviewed in Ref.~\cite{conformal}.

Since quark mass terms break the conformal symmetry of the QCD Lagrangian
explicitly one might, at first glance, expect difficulties to incorporate  
SU(3) breaking corrections into the formalism.
In fact, however, the inclusion of quark mass corrections is 
straightforward
and generates two types of effects. First, matrix elements of conformal 
operators are modified and in general do not have the symmetry of the
massless theory. This is not a ``problem'', since the conformal expansion
is designed to simplify the transverse momentum  
dependence of the wave functions by relating it to the scale dependence 
of the relevant operators. This dependence is given by operator anomalous
dimensions which are not affected by quark masses, provided they are
smaller than the scales involved. Second, new higher twist operators arise, 
in which quark masses multiply operators of lower twist. These
additional operators generate contributions to higher twist DAs, which
have been discussed in \cite{BBKT}  and are irrelevant for the
present investigation.

The conformal expansion of DAs is especially simple
when each constituent field has a fixed (Lorentz) spin projection onto the 
light-cone.
In this case, its conformal spin is
\begin{equation}
  j=\frac{1}{2}\,(l+s),
\label{eq:cspin}
\end{equation}
where $l$ is the canonical dimension  and $s$ the 
(Lorentz) spin projection.
Multi-particle states built of constituent
fields can be expanded in increasing 
conformal spin: the lowest possible spin equals the sum of spins of the
constituents, and its ``wave function'' is given by the product of one-particle
states. This state is nondegenerate and cannot mix with other states 
because of conformal symmetry. Its evolution is given by a 
simple renormalisation group equation and 
the corresponding anomalous dimension is the smallest one in the spectrum.
This state  is  the only one to
survive in the limit $Q^2\to \infty$ and is usually referred to 
as ``asymptotic distribution amplitude''.
  
As for the leading-twist quark-antiquark 
DAs we are interested in, the partial wave expansion reads
\begin{equation}
\phi(u) = 6u\ub \sum_{n = 0}^{\infty}
a_{n} C_{n}^{3/2}(2u-1) ,
\label{eq:3ceph}
\end{equation}
where $C_{n}^{3/2}(2u-1)$ are Gegenbauer polynomials.
The dimension of  quark fields is $l=3/2$ and 
the leading twist distribution corresponds to positive spin
projection $s=+1/2$ for both the quark and the antiquark. Thus, according to
(\ref{eq:cspin}), the conformal spin of each field is $j_q=j_{\bar q} = 1$;
the asymptotic distribution amplitude  is $6u\ub$. The 
Gegenbauer polynomials correspond to contributions with higher conformal 
spin $j+n$ and are orthogonal over the weight function $6u\bar u$.

Note that $a_{0} = 1$ due to the normalisation condition 
(\ref{eq:norm}). In the limit of exactly massless quarks only terms with 
even  $n$ survive in Eq.~(\ref{eq:3ceph}) because of G-parity invariance.
The conformal expansion, however, 
can be performed at the operator level and is 
disconnected from particular symmetries of states such as G-parity. The 
expansion (\ref{eq:3ceph}) is, therefore,  valid for arbitrary $n$, and it
is precisely the odd contributions to the expansion which, being
proportional to manifestly SU(3) breaking effects like the
difference of quark masses, induce the most tangible SU(3) breaking effects. 

As mentioned before, conformal invariance implies that partial waves with
different conformal spin do not mix under renormalisation to
leading-order accuracy, which means that the Gegenbauer moments
$a_n$ in (\ref{eq:3ceph}) renormalise multiplicatively:
  \begin{equation}
    a_n(\mu) = a_n(\mu_0)
  \left(\frac{\alpha_s(\mu)}{\alpha_s(\mu_0)}\right)^{(\gamma_{(n)}-
\gamma_{(0)})/{\beta_0}}
  \label{wf1}
   \end{equation}
   with $\beta_0=11 - (2/3) n_f$. The one-loop anomalous dimensions
   are \cite{GW}
  \begin{eqnarray}
  \gamma_{(n)} =  \gamma_{(n)}^{\parallel} &=& C_F 
  \left(1-\frac{2}{(n+1)(n+2)}+4 \sum_{j=2}^{n+1} 1/j\right),
  \nonumber\\
  \gamma_{(n)}^\perp &=& C_F \left(1+4 \sum_{j=2}^{n+1} 1/j\right)\,.
  \label{eq:1loopandim} 
  \end{eqnarray}
$\gamma_{(0)}$ is the anomalous dimension of the local current and
  vanishes for vector and axialvector currents. 
Note that the DAs of the pseudoscalar $K$ and the longitudinally
polarised $K^*$ have the same anomalous dimensions. 
As the $\gamma$'s are positive and increase with $n$, the 
effect of running to large scales is to damp the effects of
higher-order Gegenbauers, so that the DAs approach their asymptotic
shape. From the above-stated scaling-properties, it is evident that a
discussion of the shape of DAs is most conveniently done in terms
of Gegenbauer moments.

\section{QCD Sum Rules for Moments}
\setcounter{equation}{0}

\subsection{QCD Sum Rules}

As QCD sum rules constitute an established method for the
calculation of sta\-tic ha\-dro\-nic matrix elements like decay constants
(as opposed to dynamical quantities like form factors), we refrain
from delving into a comprehensive
explanation of technicalities, for which we refer to the original papers
\cite{SVZ} and to recent reviews \cite{alex} instead. The key feature
of the method is the use of analyticity to relate the local
short-distance operator
product expansion (OPE) of a correlation function of two currents,
\begin{equation}\label{eq1}
\Pi = i\int d^4y e^{iqy} \langle 0 | T J_1(y) J_2(0) | 0 \rangle =
\sum_n C_n(q^2) \langle O_n\rangle\equiv \Pi^{\mbox{\scriptsize OPE}}
\end{equation}
around $y = 0$ (as opposed to a
light-cone expansion aound $y^2=0$, which is appropriate
for form factor calculations, cf.~\cite{FFs})
valid for $Q^2\equiv -q^2 \ll 0$, to its dispersion relation in terms
of hadronic contributions,
\begin{equation}\label{eq2}
\Pi = \int_0^\infty ds\, \frac{\rho(s)}{s-q^2-i0}\equiv
\Pi^{\mbox{\scriptsize had}},
\end{equation}
where $\rho(s)$ is the spectral density of the correlation function
along its physical cut. 
The OPE yields a series of local operators of increasing dimension
whose expectation values $\langle O_n\rangle$ 
in the nonperturbative (physical) vacuum are
the so-called condensates. In the sum rules analysed in this paper, we
take into account the condensates listed in Tab.~\ref{tab:cond}.
\begin{table}
\renewcommand{\arraystretch}{1.3}
\addtolength{\arraycolsep}{3pt}
$$
\begin{array}{|r@{\:=\:}l||r@{\:=\:}l|}
\hline 
\quark & (-0.24\pm0.01)^3\,\mbox{GeV}^3 & \squark & (0.8\pm0.1)\,\quark\\
\gluon & (0.012\pm0.006)\,\mbox{GeV}^4 & \multicolumn{2}{l|}{} \\
\mixed & (0.8\pm0.1)\,\mbox{GeV}^2\:\quark &  \smixed & (0.8\pm0.1)
\mixed\\\hline
\multicolumn{4}{|l|}{\overline{m}_s(2\,\mbox{GeV}) = (110\pm
20)\,\mbox{MeV}~\longleftrightarrow~ \overline{m}_s(1\,\mbox{GeV})
= (149\pm 27)\,\mbox{MeV}}\\
\multicolumn{4}{|l|}{\hskip2cm\alpha_s(1\,\mbox{GeV}) = 0.513 
~\longleftrightarrow~
\Lambda_{\mbox{\scriptsize QCD}}^{(3)\mbox{\scriptsize NLO}} = 
372\,\mbox{MeV}}\\\hline
\end{array}
$$
\renewcommand{\arraystretch}{1}
\addtolength{\arraycolsep}{-3pt}
\caption[]{Input parameters for sum rules at the
  renormalisation scale $\mu=1\,$GeV. The value of $m_s$ is obtained
  from 
  quenched lattice calculations as summarised in \cite{Wittig}.
}\label{tab:cond}
\end{table}
The representation of the correlation function in terms of hadronic
matrix elements can be written as
$$
\rho(s) = f \delta(s-m_M^2) + \rho^{\mbox{\scriptsize cont}}(s),
$$
where $m_M$ is the mass of the lowest-lying state coupling to the currents
$J_{1,2}$ and $\rho^{\mbox{\scriptsize cont}}$ pa\-ra\-me\-trises all
contributions to the correlation function apart from the
ground state. $f$, the residue of the ground state pole is the
quantity one wants to determine. A QCD sum rule that allows one to do
so is obtained by equating the representations (\ref{eq1}) and
(\ref{eq2}) and
implementing the following (model) assumptions:
\begin{itemize}
\item $\rho^{\mbox{\scriptsize cont}}$ is approximated by the
   spectral density obtained from the OPE above a certain threshold,
   i.e.\ 
  $\rho^{\mbox{\scriptsize cont}}\to
\rho^{\mbox{\scriptsize OPE}}(s) \theta(s-s_0)$ with 
   $s_0\approx (m_M+\Delta)^2$ being the continuum threshold, where 
$\Delta\sim O(\Lambda_{\mbox{\scriptsize QCD}})$ 
is an excitation
  energy to be determined within the method. This assumption relies on
  the validity of global quark-hadron duality;
\item instead of the weight-functions $1/(q^2)^n$ and $1/(s-q^2)$, one
   uses different weight-functions which are optimised to
   (exponentially) suppress 
  effects of $\rho(s)$ for large values of $s$ and at the same time
   also suppress high-dimensional condensates by factorials. This is
   achieved by Borel transforming the correlation function: 
   ${\cal B}\,1/(s-q^2) = 1/M^2\exp(-s/M^2)$. A window of viable
   values of the Borel parameter $M^2$ and the continuum threshold
   $s_0$ has to be determined within the
   method itself by looking for a maximum region of minimum sensitivity (a
   plateau) in both $M^2$ and $s_0$;
\item the OPE of $\Pi$ can be truncated after a few terms. As we shall
  see (and as is well known), this condition is fulfilled only for
  low Gegenbauer moments.
\end{itemize}
After subtraction of the integral over $\rho^{\mbox{\scriptsize
    OPE}}$ above $s_0$ from both sides, the final sum rule reads
\begin{equation}\label{eq:SR}
{\cal B}_{\mbox{\scriptsize sub}}\,\Pi^{\mbox{\scriptsize OPE}} \equiv
 \frac{1}{M^2}\int_0^{s_0} ds\,e^{-s/M^2}\,
\rho^{\mbox{\scriptsize OPE}}(s) = \frac{f}{M^2}\,e^{-m_M^2/M^2},
\end{equation}
which gives the hadronic quantity $f$ as a function of the Borel
parameter $M^2$ and the continuum threshold $s_0$ (and the condensates
and short-distance parameters from the OPE).

Even Gegenbauer moments can be determined from diagonal
correlation functions of type
\begin{equation}\label{eq:3.4}
i\int d^4y e^{iqy} \langle 0 | T \bar q(y) \Gamma s(y) \bar s(0)\Gamma
[0,z] q(z) | 0\rangle,
\end{equation}
with a suitably chosen Dirac structure $\Gamma$. This form of the
 correlation function differs from Refs.~\cite{CZreport,BB96}, where
 local operators with an arbitrary number
 of covariant derivatives were used rather than the nonlocal
 operator $\bar s(0)\Gamma [0,z] q(z)$. We find the calculation with
 nonlocal operators very convenient, as it allows one to calculate
 all moments in one go. 
Specifying for instance to $\Kpar$, the sum rule reads
\begin{equation}\label{eq:x}
{\cal B_{\mbox{\scriptsize sub}}}\,\Pi^{\mbox{\scriptsize OPE}} =
(f_K^\parallel)^2\,e^{-m_{K^*}^2/M^2}\,\frac{1}{M^2}\int_0^1
du\,e^{i\ub\qz}\phi_K^\parallel(u),
\end{equation}
where also $\Pi^{\mbox{\scriptsize OPE}}$ is expressed as integral
over $u$, which naturally emerges as Feynman parameter in the
calculation, and comes with the same weight function $\exp(i\ub\qz)$.
Sum rules for individual 
Gegenbauer moments are obtained by expanding both sides in powers
of $\qz$, or effectively replacing
$$e^{i\ub\qz}\to C_n^{3/2}(2u-1),\qquad \int_0^1
du\,e^{i\ub\qz}\phi_K^\parallel(u)\to
\frac{3(n+1)(n+2)}{2(2n+3)}\,a_n^\parallel.$$
For odd moments, one analyses
 nondiagonal correlation functions
 of type
\begin{equation}
i\int d^4y e^{iqy} \langle 0 | T \bar q(y) \Gamma_2 s(y) \bar s(0)\Gamma_1
[0,z] q(z) | 0\rangle
\end{equation}
with structures $\Gamma_1$ and $\Gamma_2$ of opposite chirality
\cite{Russians,CZreport}.
The $\Gamma_i$ appropriate for $K$, $\Kpar$ and $\Kperp$ are given in
App.~\ref{app:a}, where we also give complete expressions for
perturbative and quark-condensate
contributions to $O(\alpha_s)$, as well as tree-expressions for 
the dimension 5 mixed condensate.

Formulas for the correlation functions and Gegenbauer moments are
collected in the appendices. Note in particular the scaling of
different contributions to the Gegenbauer moments in $n$, the order of
the moment: nonperturbative terms increase with positive powers of
$n$ with respect to the perturbative contribution. For large $n$, 
this behaviour upsets the usual hierarchy of contributions to the
OPE, where nonperturbative terms are expected to be a moderately sized 
correction to the
leading term. The origin of this behaviour can
be easily understood from the fact that in the local expansion the
vacuum fields have exactly zero momentum, which
yields the $\delta$-function terms in (\ref{eq:cond}) and
(\ref{eq:cond2}). This amounts to a multipole expansion of the DA
around its endpoints, which is justified if one is only interested in gross
features of the DA, like the first few moments, but which is clearly
inappropriate for extracting more detailed information on the shape. In the 
present paper we adopt the viewpoint that at least the first two
Gegenbauer moments can be reliably obtained from local sum rules of
the type discussed above.

\subsection{A Short Discussion of Decay Constants and Lattice Results}

\begin{figure}[tb]
$$\epsfysize=0.23\textheight\epsffile{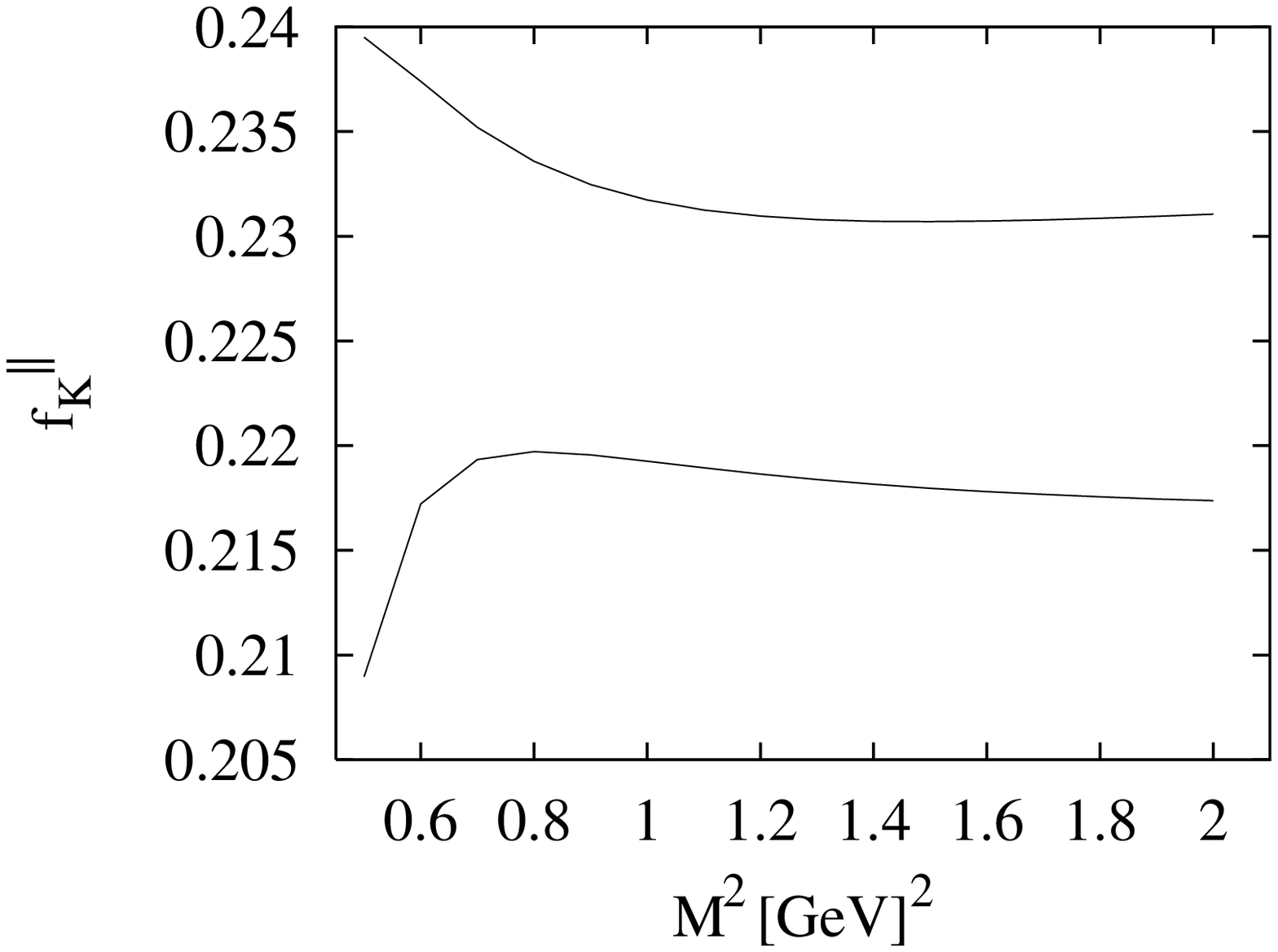}
$$
\caption[]{$f_K^\parallel$ from QCD sum rules as function of the Borel
  parameter. The spread between the
  two curves corresponds to the uncertainty induced by varying the
  input parameters within their error margins.}\label{fig:fL}
$$\epsfysize=0.23\textheight\epsffile{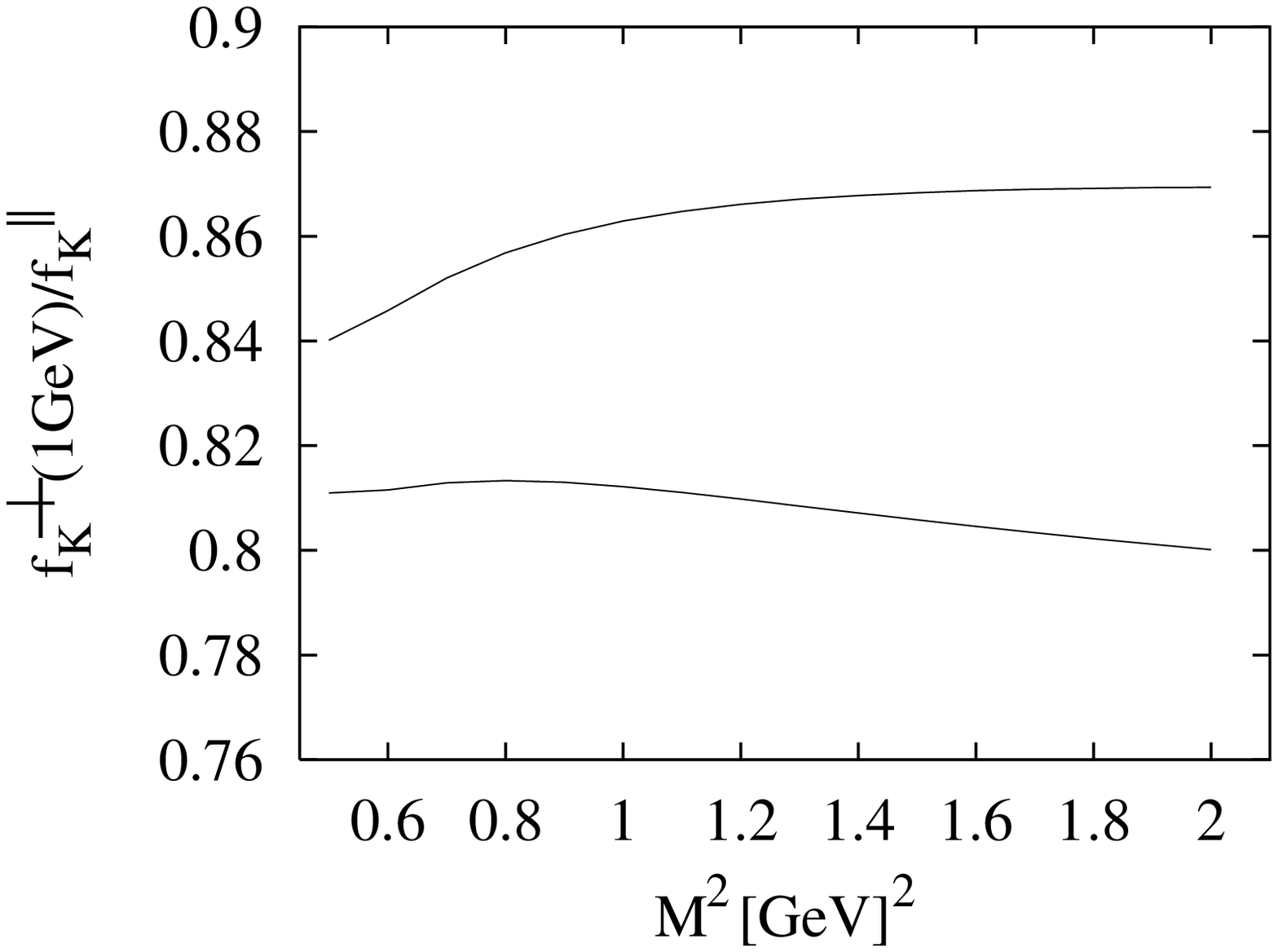}
$$
\caption[]{The ratio $f_K^\perp(1\,{\rm GeV})/f_K^\parallel$ as
  function of the Borel parameter $M^2$. Interpretation of curves as
  in previous figure. $s_0$
  is varied between 1.1 and 1.3$\,$GeV$^2$ for $f_K^\perp$
  and 1.7 and 1.9$\,\mbox{GeV}^2$ for $f_K^\parallel$.}\label{fig:decay}
\end{figure}

The decay constant of the pseudoscalar K meson is well known from
$K^+\to \mu^+\nu_\mu (+\gamma)$ and quoted as \cite{PDG} $$f_K = (159.8\pm
1.3)\,\mbox{MeV}.$$  The decay constant $f_K^\parallel$ can be
extracted from the branching ratio of $\tau^-\to K^{*-}\nu_\tau$ via
$$
B(\tau^-\to K^{*-}\nu_\tau) = \frac{G_F^2m_\tau |V_{us}|^2}{8\pi}\,
\tau_\tau m_{K^*}^2 (f_K^\parallel)^2\left( 1 +
\frac{m_\tau^2}{2m_{K^*}^2} \right) \left( 1 -
\frac{m_{K^*}^2}{m_\tau^2} \right)^2.
$$
With $|V_{us}|= 0.220$ and the other parameters taken from \cite{PDG}, one
finds
$$f_K^\parallel = (217\pm 5)\, {\rm MeV}.$$
It is interesting to compare this value with a ``postdiction'' from
QCD sum rules. We use the sum rule quoted as (C.4) in \cite{BBKT} with
the input parameters as in Tab.~\ref{tab:cond}. From the result
plotted in Fig.~\ref{fig:fL} we conclude
$$
f_K^\parallel{}^{\mbox{\scriptsize (SR)}} = (225\pm 7)\,\mbox{MeV}
$$
{}and $s_0 = (1.8\pm0.1)\,$GeV$^2$. The plateau in $M^2$ extends from
1 to 2~GeV$^2$. Although the error only includes the
impact of varying the input parameters of Tab.~\ref{tab:cond} within
their respective ranges and does not account for any systematic
uncertainties, the agreement with the experimental result is
remarkably good.

The value of the third constant, $f_K^\perp$, has not been measured
yet, but has been calculated in quenched lattice QCD and from QCD sum rules. 
The lattice result for the ratio of the $K^*$ decay constants is
\cite{tensorlattice}
$$
\frac{f_K^\perp(2\,\mbox{GeV})}{f_K^\parallel} = 0.739(17).
$$
{}From QCD sum rules, on the other hand, we find, cf.\ Fig.~\ref{fig:decay},
$$
\frac{f_K^\perp(1\,\mbox{GeV})}{f_K^\parallel} = 0.84(03)
\quad\longleftrightarrow\quad
\frac{f_K^\perp(2\,\mbox{GeV})}{f_K^\parallel} = 0.74(03),
$$
using the formulas (C.4) and (C.5) in \cite{BBKT}, which include NLO
radiative corrections. The optimum continuum threshold for $f_K^\perp$ 
turns out to be $s_0=(1.2\pm 0.1)\,$GeV$^2$. The fact that $s_0$ for
$\Kperp$ is smaller than for $\Kpar$ is in agreement with the findings
of Ref.~\cite{BB96} and related to the fact that, with the Dirac structure
$\Gamma$ chosen as specified in App.~\ref{app:b}, the sum rule does
not only receive contributions from $K^*$, but also from the ground
state in the opposite parity channel, the 1$^+$ state $K_1(1270)$. This
contribution has to be included in the continuum, which results in a
low value of $s_0$. Note that this is not an external condition
imposed by us when evaluating the sum rule, but emerges naturally from
the criterion of stability and the presence of a plateau in $M^2$. 
We consider the ratio of decay constants rather
than the constants themselves, as effects of unknown higher order
corrections are expected to cancel in the ratio. The sum rules are
evaluated at the low scale 1 GeV and the scaling from 1 to 2~GeV is
done using NLO renormalisation group improvement. 
Note that the dependence on the Borel
parameter largely cancels in the ratio of decay constants and also the
dependence on $s_0$ is rather mild. The agreement with the lattice
results is remarkably good. Using the experimental value for
$f_K^\parallel$, QCD sum rules hence predict, to NLO accuracy and
including 2-loop running,
\begin{equation}\label{update}
\begin{array}[b]{r@{\qquad}l}
f_K^\perp(1\,\mbox{GeV}) = (182\pm 10)\,{\rm MeV}, &
f_K^\perp(2\,\mbox{GeV}) = (160\pm 9)\,{\rm MeV},\\[10pt]
f_K^\perp(4.8\,\mbox{GeV}) = (149\pm 8)\,{\rm MeV}.
\end{array}
\end{equation}
This updates the result obtained in Ref.~\cite{BBKT}.

\subsection{\boldmath The Gegenbauer Moment $a_1$}

\begin{figure}[p]
$$\epsfysize=0.25\textheight\epsffile{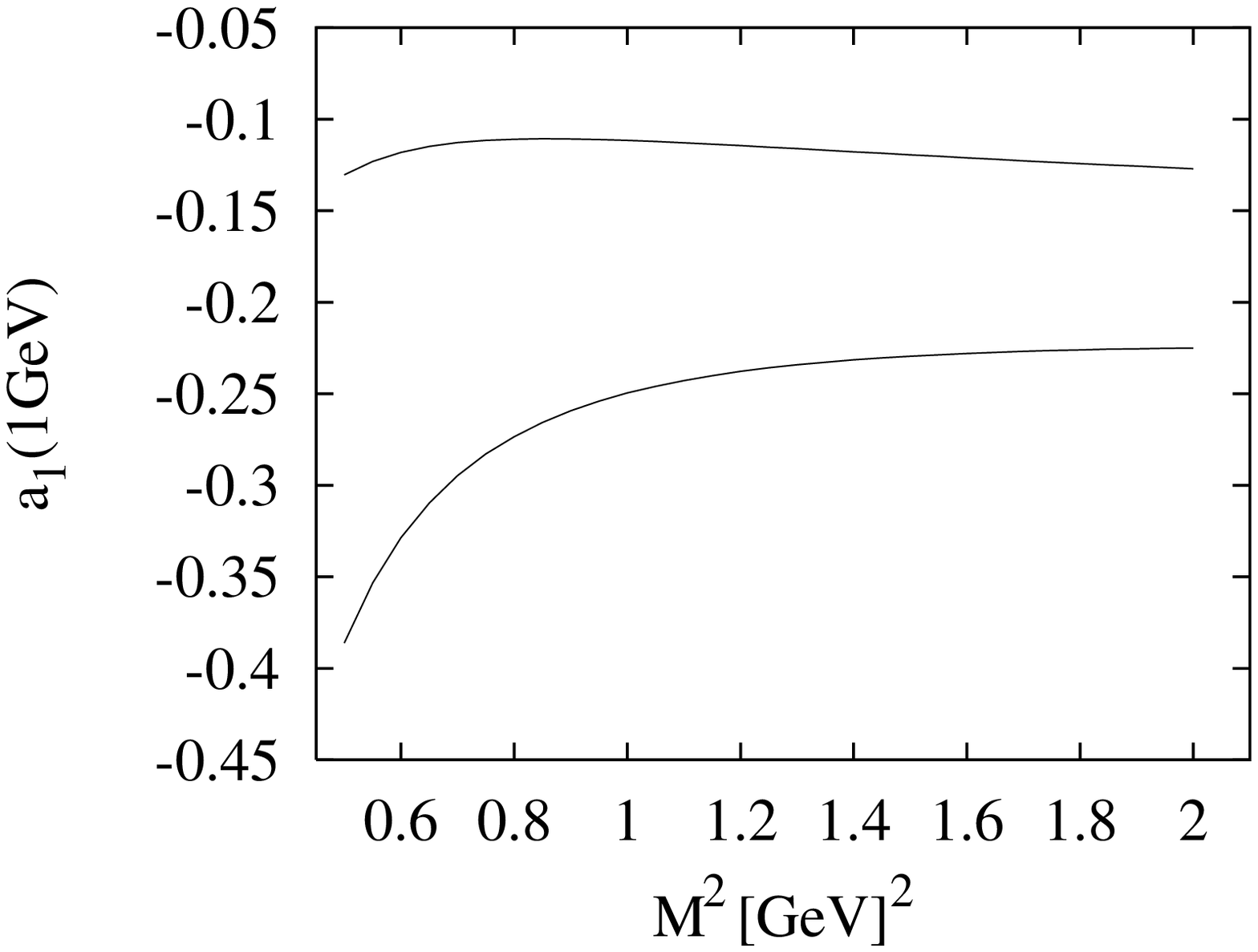}$$
\caption[]{$a_1(1\mbox{GeV})$ as function of the Borel parameter
  $M^2$.  The spread between the two curves corresponds to the 
uncertainty induced by varying the input parameters within their 
error margins. $s_0$ is fixed at 1.8~GeV$^2$.}\label{a1K}
$$\epsfysize=0.25\textheight\epsffile{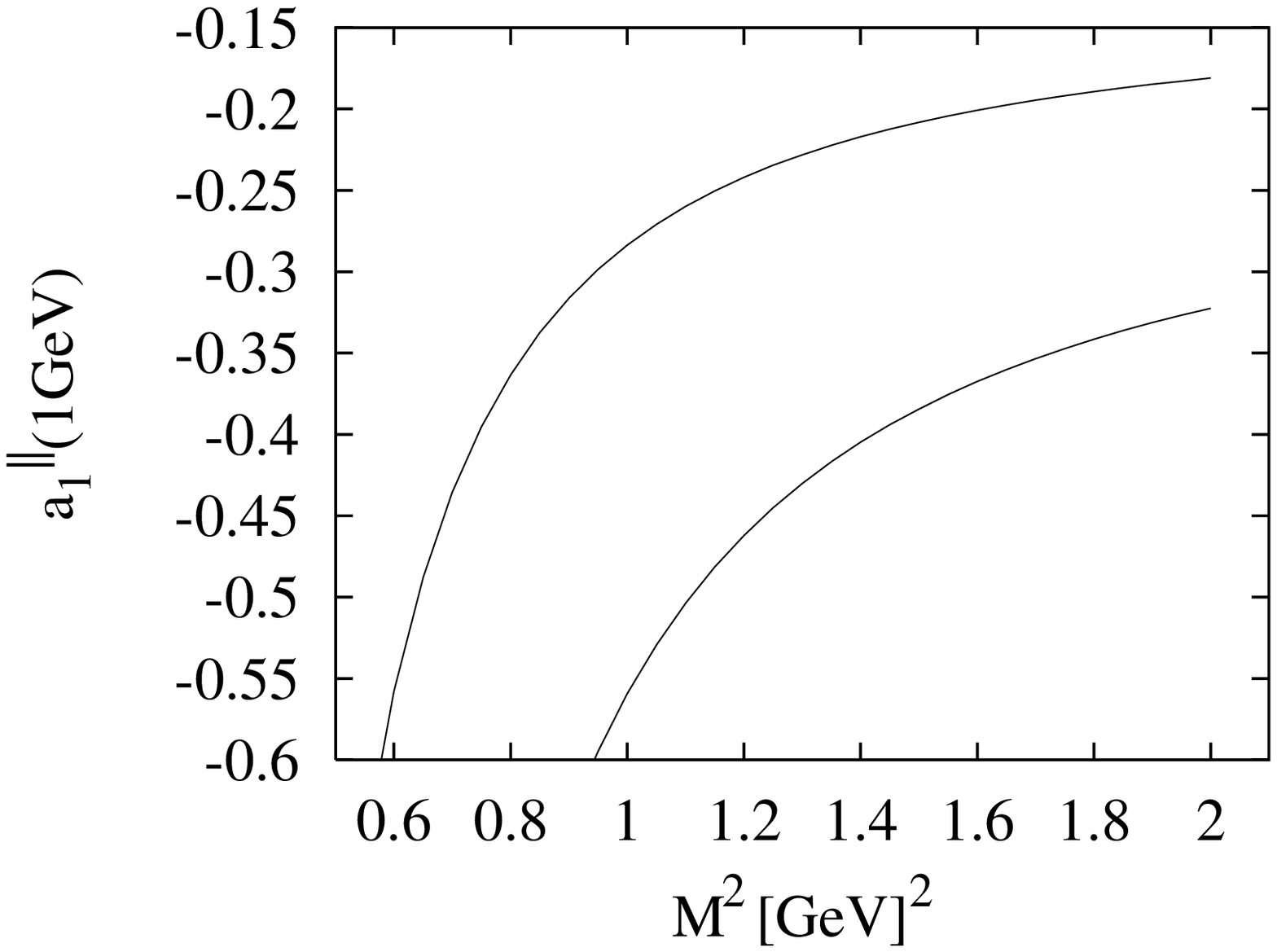}$$
\caption[]{$a_1^\parallel(1\mbox{GeV})$; notations and conventions as
  above.}\label{a1par}
$$\epsfysize=0.25\textheight\epsffile{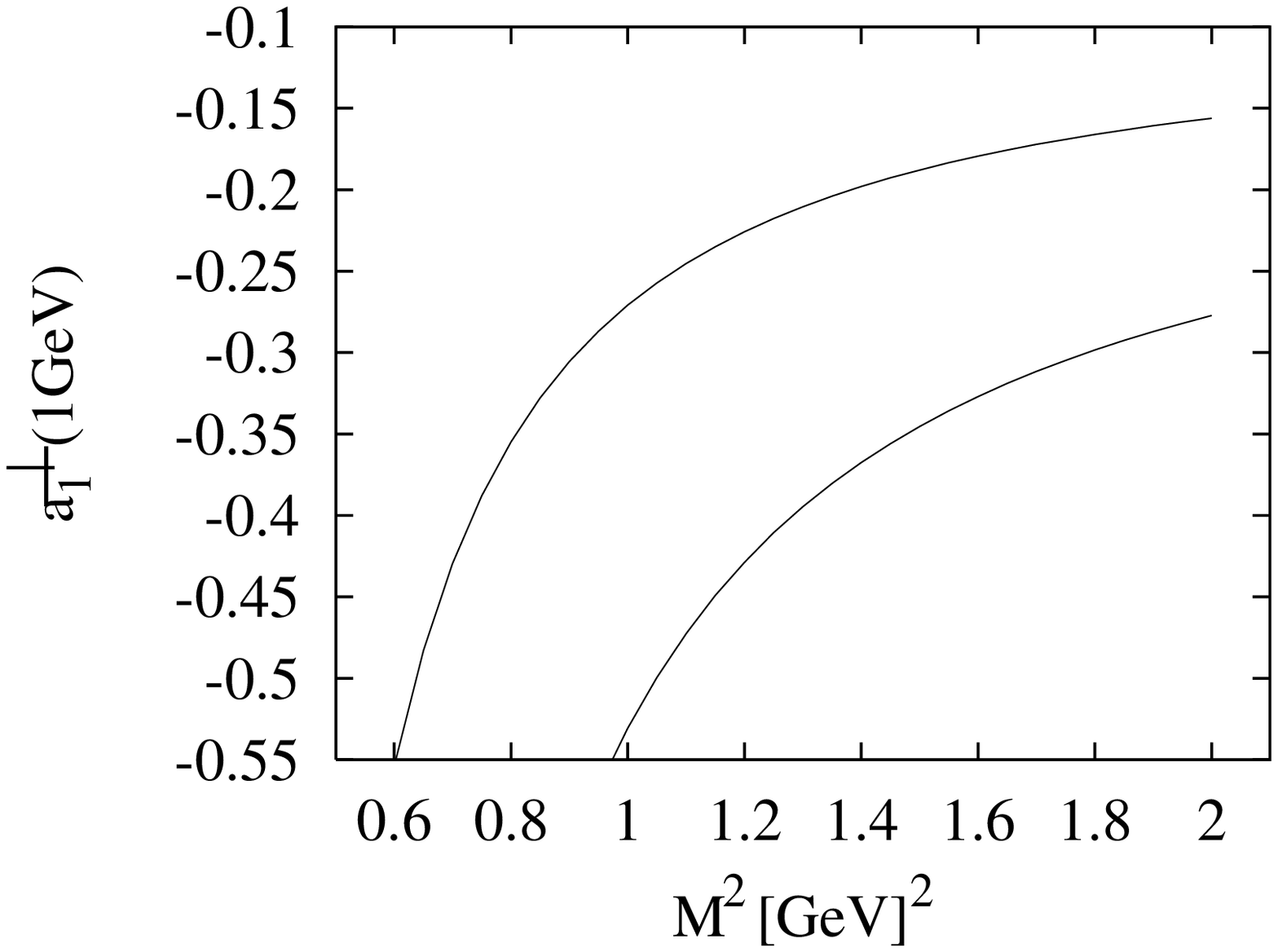}$$
\caption[]{$a_1^\perp(1\mbox{GeV})$; $s_0$ is fixed at 1.2~GeV$^2$; other
notations and conventions as  above.}\label{a1perp}
\end{figure}

Let us now calculate the first moments of the $K$ and
$K^*$ DAs. Previous determinations go back to Refs.~\cite{CZreport,BBKT},
where the following values were obtained:
$$
a_1(1\,\mbox{GeV}) = 0.17,\quad a_1^\parallel(1\,\mbox{GeV}) = 0.19\pm
0.05, \quad a_1^\perp(1\,\mbox{GeV}) = 0.20\pm 0.05.
$$
These results are valid at LO in QCD and were obtained using
QCD sum rules which, unfortunately, 
partly suffered from mistakes. In particular, we
find a different {\em sign} of the perturbative contribution w.r.t.\
the formulas given in Ref.~\cite{CZreport,BBKT}. 
We have  carefully checked that the sign we obtain is
indeed the correct one: for the $n$th Gegenbauer moment, the leading
order perturbative contribution is the same for all three correlation
functions and is given by
$$
 \frac{3}{4\pi^2}\, m_s \frac{1}{M^2}\, \left( 1 - e^{-s_0/M^2}\right)
 \int_0^1 C_n^{3/2}(2u-1) \ub = \frac{3}{4\pi^2}\, m_s \frac{1}{M^2}\,
 \left( 1 - e^{-s_0/M^2}\right) \left(-\frac{1}{2}\right)^n.
$$
It appears that the factor $(-1)^n$ has been missed in \cite{CZreport}.
Due to the change in sign of the perturbative
contribution, the leading order sum rule becomes numerically unstable: 
perturbation theory and the mixed condensate yield negative
contributions, the quark condensate a positive one, and the sum is 
close to zero.
This is in stark contrast to the statement made in
Ref.~\cite{CZreport}, according to which the sum rule is dominated by the
quark condensate contribution and yields a positive result for all 
$a_1$.

The different signs of individual contributions, together with the
fact that none of them is numerically dominant, entails that no 
meaningful result can be extracted from the leading order sum rule. The
situation improves, however, if one includes radiative corrections.   
We have calculated $O(\alpha_s)$ corrections to both the
perturbative and the quark condensate contribution and find that they
reduce the size of the quark condensate contribution, but increase
that of the
perturbative contribution. As a result, the perturbative contribution
becomes numerically dominant, as is actually expected for 
a ``good'' sum rule. 
By varying all the input parameters
of Tab.~\ref{tab:cond} within their respective ranges we obtain
\begin{equation}\label{resultsa1}
a_1(1\,\mbox{GeV}) = -0.18\pm 0.09, \quad a_1^\parallel(1\,\mbox{GeV})
= -0.4\pm 0.2, \quad  a_1^\perp(1\,\mbox{GeV})
= -0.34\pm 0.18.
\end{equation}
The Borel window is taken to be 1 to 2$\,\mbox{GeV}^2$, as
motivated by the results of the previous subsection. Note that the
l.h.s.\  of the sum rules for $a_1^{\parallel,\perp}$ 
contains the factor $f_K^\parallel f_K^\perp$, which we substitute by 
their respective sum rules instead of using the values determined
in the previous subsection, the reason being that one expects unknown
higher order corrections to cancel in the ratio.
The dependence of the individual sum rules on the input
parameters is plotted in
Figs.~\ref{a1K}, \ref{a1par} and \ref{a1perp}. The error bars in
(\ref{resultsa1}) are rather large, which is due to the fact that the
first Gegenbauer moments are explicitly proportional to SU(3) breaking
quantities, i.e.\ $m_s$ or the difference of condensates, e.g.\ $\quark
- \squark$, which come with a considerable uncertainty. 

\begin{figure}
$$\epsfysize=0.23\textheight\epsffile{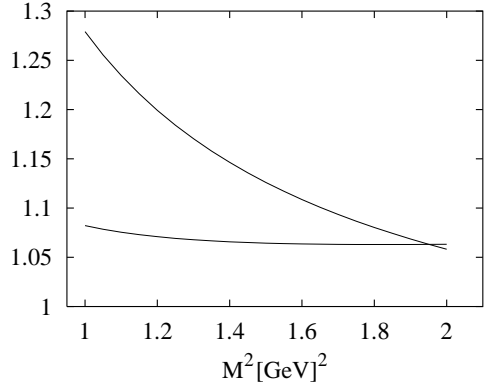}$$
\caption[]{$a_0$ from nondiagonal sum rules for $K$ (lower curve) and
  $K^*$ (upper curve) for central values of input parameters.}\label{fig:a0}
\end{figure}

We would also like to add that the use of nondiagonal sum rules has
been met with criticism. It has been argued in Ref.~\cite{BBK}
that these sum rules may suffer from large
contributions of higher resonances or, in the case of the $K$, from
instanton contributions to the pseudoscalar current. 
We can actually estimate the amount of possible contamination of these
sum rules by studying their local limit,
yielding $a_0$, which is 1 by definition. In Fig.~\ref{fig:a0} 
we plot the sum rules for
$a_0$ for both the $K$ and the $K^*$ (there is only one sum rule for
the $K^*$), for central values of the input
parameters, $s_0 = 1.2\,{\rm GeV}^2$ for $K^*$, $s_0 = 1.8\,{\rm
  GeV}^2$ for the $K$, and replacing $f_K^{\perp,\parallel}$ in the
denominator by their respective sum rules.
{}The plot shows no sizeable contamination for the $K$, and a
moderate one for the $K^*$; both results lie in the ball-park of the
expected accuracy of QCD sum rules and come with uncertainties from
the input parameters,  in particular for $a_0$ of the
$K$, which is directly proportional to $m_s$. Radiative corrections
are important in both cases and drag the result closer to
$a_0=1$. This result strengthens our
confidence in the suitability of nondiagonal sum rules for extracting
meaningful values of $a_1$.

\subsection{\boldmath The Gegenbauer Moment $a_2$}

Sum rules for even moments and the effects of SU(3) breaking have been
studied in various papers, e.g.\
\cite{Russians,CZreport,BBKT}. None of these papers, however, does
contain a complete set of sum rules for Gegenbauer moments. Moreover, 
while checking our
two-loop and nonperturbative calculations against the formulas
collected in Ref.~\cite{BBKT}, we found that 
the terms in $m_s\smixed$, given correctly in Ref.~\cite{CZreport}
  for the moments of $K^{(*)}$, have not been correctly translated
  into Gegenbauer moments in Ref.~\cite{BBKT}; we have calculated
  these contributions anew for all three DAs, using nonlocal
  currents, and confirm the results of Ref.~\cite{CZreport}.
We have also recalculated the four-quark contributions and confirm the
expressions quoted in \cite{BBKT}. The complete set of (hopefully)
correct sum rules is given in App.~\ref{app:b}.

With the input parameters from Tab.~\ref{tab:cond} we obtain
\begin{equation}\label{resultsa2}
a_2(1\,\mbox{GeV}) = 0.16\pm 0.10, \quad a_2^\parallel(1\,\mbox{GeV})
= 0.09\pm 0.05, \quad  a_2^\perp(1\,\mbox{GeV})
= 0.13\pm 0.08
\end{equation}
by varying all the input parameters of
Tab.~\ref{tab:cond} within their respective
ranges. 
These numbers have to be compared with those quoted in the first
reference in \cite{FFs} and in Ref.~\cite{BBKT}:
$$
a_2(1\,\mbox{GeV}) = 0.2,\quad a_2^\parallel(1\,\mbox{GeV})
= 0.06\pm 0.06, \quad  a_2^\perp(1\,\mbox{GeV})
= 0.04\pm 0.04.
$$
The discrepancy to (\ref{resultsa2}) 
is mainly due to the wrong expressions for the
contribution of the $\smixed$ condensate used in \cite{BBKT}. 

\begin{figure}[p]
$$\epsfysize=0.25\textheight\epsffile{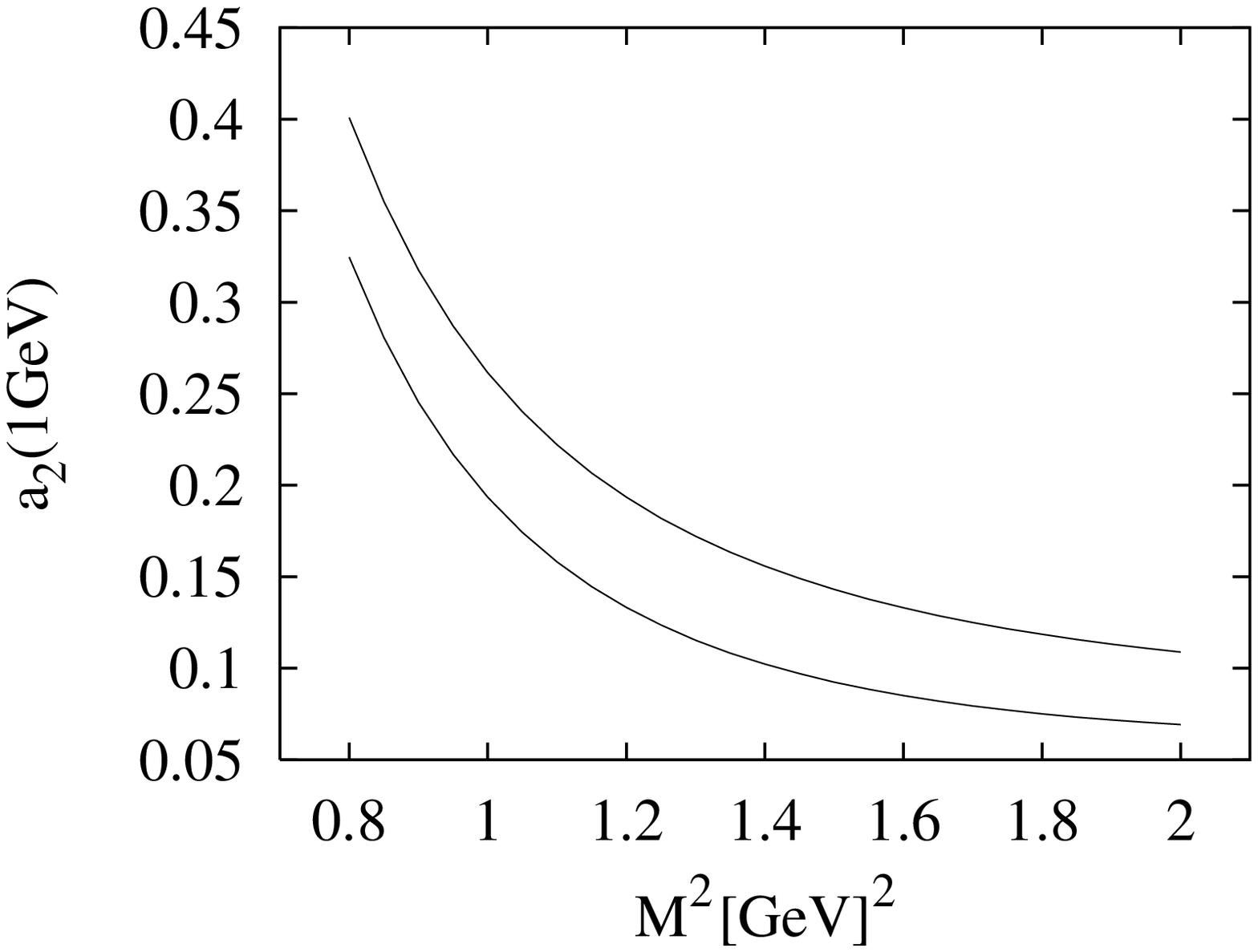}
$$
\caption[]{$a_2(1\mbox{GeV})$ as function of the Borel parameter
  $M^2$. The spread between the two curves corresponds to the 
uncertainty induced by varying the input parameters within their 
error margins. $s_0$ is fixed at 1.8~GeV$^2$.}\label{a2K}
$$\epsfysize=0.25\textheight\epsffile{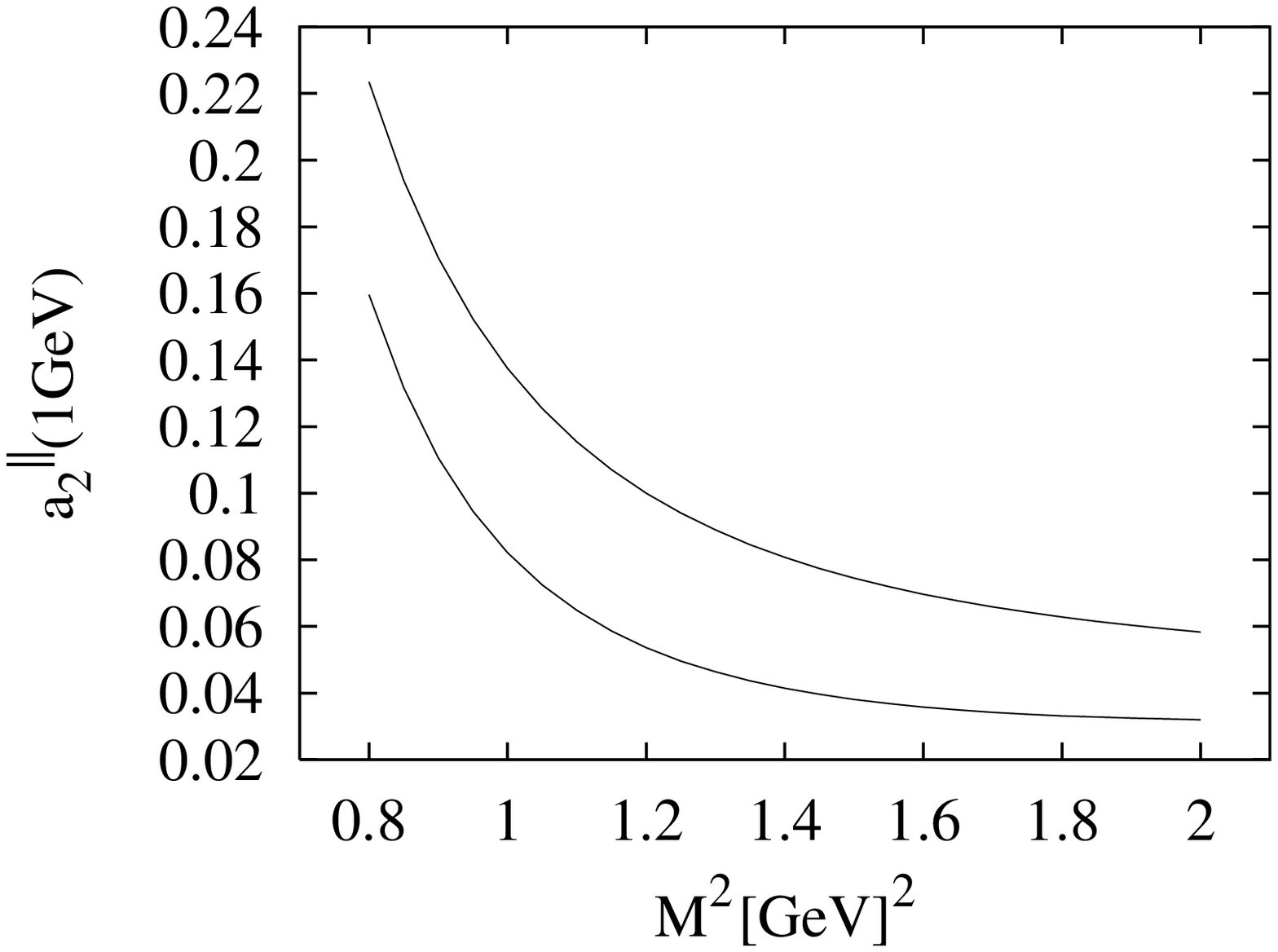}
$$
\caption[]{$a_2^\parallel(1\mbox{GeV})$; notations and conventions as
  above.}\label{a2par}
$$\epsfysize=0.25\textheight\epsffile{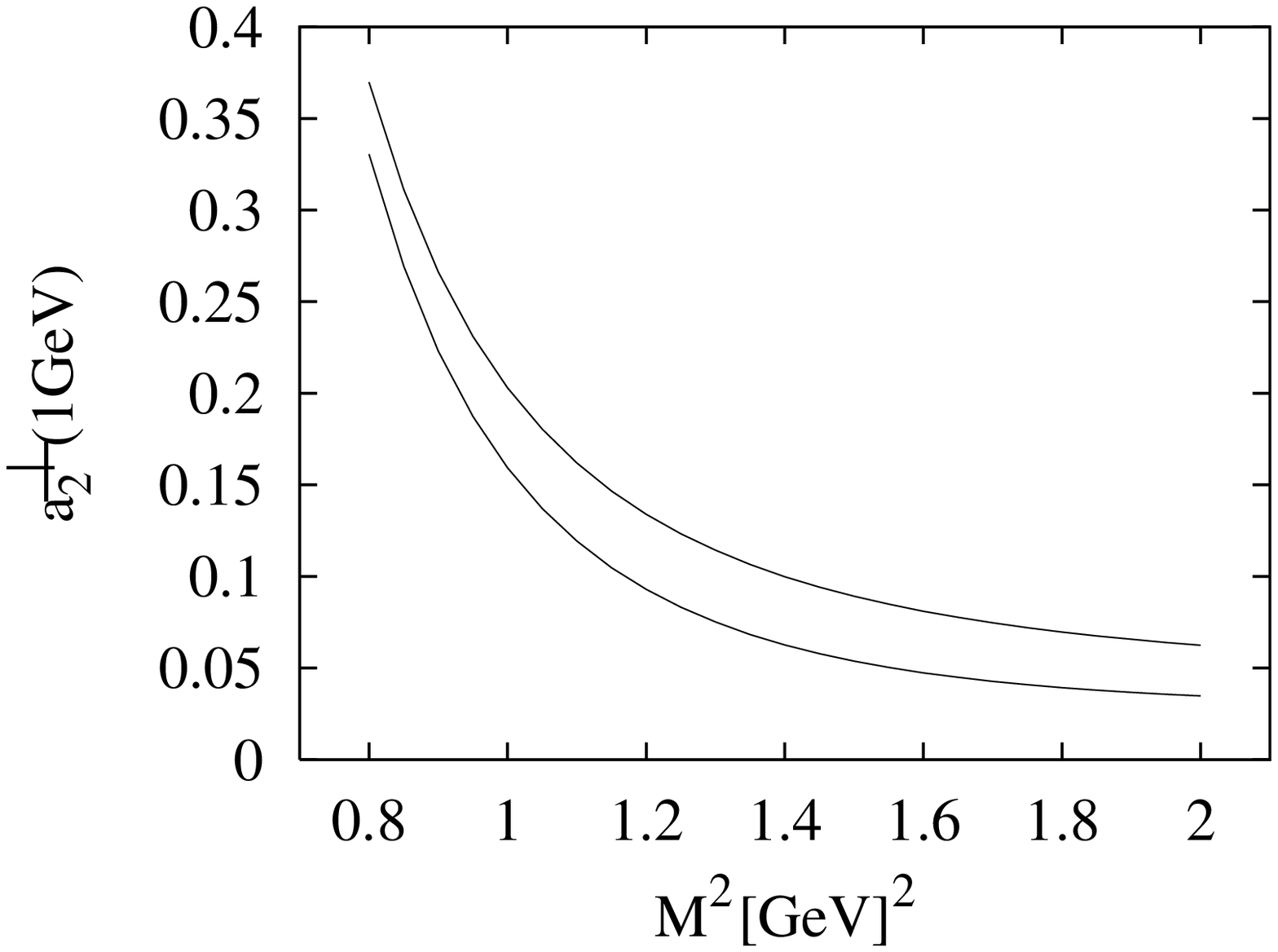}
$$
\caption[]{$a_2^\perp(1\mbox{GeV})$;  $s_0$ is fixed at 1.2~GeV$^2$; other
notations and conventions as
  above.}\label{a2perp}
\end{figure}

Like for $a_1$, the relative errors of the moments (\ref{resultsa2}) are
considerably larger than the ones quoted for the decay constants. The
reason is the absence of a stability plateau in $M^2$ as the
perturbative contribution is of $O(\alpha_s)$ and small. For the
extraction of the Gegenbauer moments we thus have to rely
on the optimum values of $M^2$, 1 to 2$\,\mbox{GeV}^2$, 
and $s_0$ determined from the sum rules
for the decay constants.

\subsection{Models for Distribution Amplitudes}

Nearly all models for DAs rely on a truncated conformal expansion,
which is not too surprising in view of the fact that it is only
moments (or Gegenbauer moments) that are accessible to direct
calculation to date. Truncation
after the first few moments implies that only gross features of
the DAs are taken into account, which is sufficient if physical
amplitudes are obtained by convoluting with a 
  smooth perturbative scattering amplitude -- or if the
characteristic scale of the process is large enough for the
logarithmic damping of high-order Gegenbauer moments to have become
effective. In this paper we assume that this is indeed the case and
that a truncation after $n=2$ yields a meaningful approximation. We
then obtain  the following model for the three
leading-twist DAs of strange mesons containing an s quark:
\begin{equation}
\phi_K(u,\mu) =  6u(1-u)\left[ 1 + 3 a_1(\mu) (2u-1) +
  a_2(\mu)\,\frac{3}{2}\,( 5(2u-1)^2 - 1)\right].
\label{eq:forthesimpleminded}
\end{equation}
To obtain the DAs for mesons with an s antiquark, one has to replace
$u\leftrightarrow 1-u$. The values of the Gegenbauer moments $a_1$ and
$a_2$ are calculated from QCD sum rules at the low
scale 1 GeV and scaled up to 2 and 4.8~GeV, to LO
accuracy;\footnote{NLO scaling mixes all Gegenbauer moments, so that
to this accuracy 
the model (\ref{eq:forthesimpleminded}) receives radiatively generated
corrections in $a_{n>2}$.} numerical values are collected in
Tab.~\ref{tab:2}. Eq.~(\ref{eq:forthesimpleminded}) is ready for use 
in phenomenological applications.
\begin{table}
\renewcommand{\arraystretch}{1.2}
\addtolength{\arraycolsep}{3pt}
$$
\begin{array}{|l|rrr|}
\hline
& \multicolumn{1}{c}{K} &  \multicolumn{1}{c}{\Kpar} & 
\multicolumn{1}{c|}{\Kperp}\\\hline 
a_1 (1\,{\rm GeV}) & -0.18\pm0.09& -0.40\pm0.20 & -0.34\pm0.18 \\ 
a_1 (2\,{\rm GeV}) & -0.15\pm0.07& -0.32\pm0.16& -0.28\pm0.14 \\
a_1 (4.8\,{\rm GeV}) & -0.13\pm0.06& -0.28\pm0.14 & -0.23\pm0.11\\\hline
a_2 (1\,{\rm GeV}) & 0.16\pm 0.10 & 0.09\pm 0.05& 0.13\pm 0.08 \\
a_2 (2\,{\rm GeV}) & 0.11\pm0.07 & 0.06\pm0.04 & 0.10\pm0.06\\
a_2 (4.8\,{\rm GeV}) &0.09\pm0.06 & 0.05\pm0.03& 0.08\pm0.05\\\hline
\end{array}
$$
\renewcommand{\arraystretch}{1}
\addtolength{\arraycolsep}{-3pt}
\caption[]{First and second Gegenbauer moments $a_1$ and $a_2$ 
of the leading-twist DAs of $K$ and $K^*$ mesons evaluated at 
different scales.}\label{tab:2}
\end{table}

\section{Summary and Conclusions}

We have presented a comprehensive study of 
SU(3) breaking corrections to the decay constants and the 
leading-twist light-cone
distribution amplitudes of $K$ and $K^*$ mesons from QCD sum rules. We
have corrected a few computational mistakes in previous works and
included radiative corrections to the leading contributions to 
odd moments of the distribution
amplitudes. Our main results are summarised in
Eqs.~(\ref{update}), (\ref{eq:forthesimpleminded}) and Tab.~\ref{tab:2}. 
We find in particular that the first Gegenbauer moments are {\em
  negative} for all three DAs, in contrast to previous determinations.
The decay constants $f_K$ and $f_K^\parallel$ can be extracted from
experiment with good accuracy, whereas $f_K^\perp$ has to be
determined theoretically. Its accuracy  is
limited mainly by the uncertainty of the input parameters, whereas the
accuracy of
the Gegenbauer moment $a_1$ is in addition limited by the lack of a proper
plateau region in the Borel parameter, which suggests that the uncertainty
is dominated by systematics. The latter limitation also affects $a_2$.
This implies that further refinement of
the sum rules by calculating even higher order corrections is not
likely to improve their accuracy.
Any decisive improvement has to come from a
different method, for which lattice QCD appears to be a natural
candidate. It is to be hoped that the preliminary results
for the second moment of the twist-2 pion DA reported in
\cite{latticeDAs} will be improved and also extended to the $K$ and
other mesons.

Eq.~(\ref{eq:forthesimpleminded}) and Tab.~\ref{tab:2} present an
approximation to the full DA to NLO in the conformal expansion. A
question we have not touched upon in this paper is the suitability of
this approximation for calculating physical amplitudes of type
$$
A_{\mbox{\scriptsize phys}} \sim \int_0^1 du\,\phi(u) T(u),
$$
where $T(u)$ is a perturbative scattering amplitude. Obviously the
answer to this question depends on the functional dependence of $T$
on $u$. The standard argument to justify the validity of a truncated
conformal expansion is that for sufficiently smooth functions $T(u)$
higher order Gegenbauer polynomials in $\phi(u)$ with their highly
oscillatory behaviour are effectively washed out. It is also argued
that, for large scales, higher order Gegenbauer moments are suppressed
by renormalisation group scaling, so that for most applications the
asymptotic DA is  sufficient. These arguments are usually
presented in a qualitative rather than a quantitative way and it would
be interesting to study, within a well-defined model that does not
rely on conformal expansion, how sensitive physical amplitudes
actually are to a truncation of the conformal expansion, 
and at what scales logarithmic damping becomes
effective. These
questions will be addressed in a future publication.

\subsection*{Acknowledgements}
We are grateful to V.~Sanz for collaboration in the early stages of
this work.

\appendix
\renewcommand{\theequation}{\Alph{section}.\arabic{equation}}
\setcounter{table}{0}
\renewcommand{\thetable}{\Alph{table}}

\section*{Appendix}
\section{Results for Nondiagonal Correlation Functions}
\label{app:a}
\setcounter{equation}{0}

We calculate the correlation function
\begin{equation}\label{eq:CF}
\Pi_{M} = i\int d^4y e^{iqy} \langle 0 | T \bar q(y)
\Gamma_2 s(y)\: \bar s(0) \Gamma_1 [0,z] q(z) | 0 \rangle,
\end{equation}
where $\Gamma_{1,2}$ are Lorentz-structures suitable for the
projection onto the meson $M$, $M\in\{ K,\,\Kperp,\,\Kpar\}$, $z^2=0$
is a light-like vector and 
$$
[0,z] = P\exp\left\{ -ig \int\limits_0^1 dt\, z_\mu A^\mu(tz)\right\}
$$
the path-ordered gauge-link joining the quark fields in the bilinear $\bar
s(0)\Gamma_1 q(z)$. The above sign-convention for $g$ implies that the
covariant derivative is given by $D_\mu = \partial_\mu - i g A_\mu$. Odd
Gegenbauer moments are most conveniently determined from 
nondiagonal correlation functions, involving one chiral-odd and one
chiral-even current. To be specific, we choose the currents given in
Tab.~\ref{tab:A} and perform an operator product expansion of $\Pi_M$
up to dimension 5 condensates.
\begin{table}
\renewcommand{\arraystretch}{1.3}\addtolength{\arraycolsep}{4pt}
$$
\begin{array}{|l|lll|}
\hline
& \Gamma_1 & \Gamma_2 & \mbox{projection~}{\cal P}: \Pi_M\to {\cal
  P}\Pi_M\\\hline
K & \hat{z}\gamma_5 & \gamma_5 & -1/\qz\\
\Kpar & \hat{z} &  \sigma_{\mu\nu}q^\mu z^\nu & i/\qz^2\\[5pt]
\Kperp & \sigma_{\mu\nu} & \gamma_\alpha &\displaystyle
\frac{1}{D-2}\,\frac{-i}{\qz^2} \left\{ \qz g^{\mu\alpha} z^\nu - q^\mu
z^\nu z^\alpha \right\}\\[10pt]\hline
\end{array}
$$
\renewcommand{\arraystretch}{1}\addtolength{\arraycolsep}{-4pt}
\caption[]{Dirac structures used in Eq.~(\ref{eq:CF}). The choice of
  $\Gamma_{1,2}$ together with the projection $\cal P$ projects onto
  the twist-2 structure and eliminates  factors $i\qz$ picked up by
  the traces. $D$
  is the number of dimensions in dimensional regularisation. We
  use the notation $\hat{a} = a_\mu \gamma^\mu$ for an arbitrary
  4-vector $a$.}\label{tab:A}
\end{table}
We express the perturbative results in the form 
$$
{\cal P}\Pi_M^{\mbox{\scriptsize pert.th}} = - \frac{3}{4\pi^2}\, 
\ms \ln \,\frac{-q^2}{\mu^2}\, \int\limits_0^1
e^{i\bar u \qz}\, \bar u\, \pi_M(u;\mu) + \mbox{~terms analytic in $q^2$},
$$
where $u$ is the momentum fraction carried by the s quark in the $K$
or $K^*$ meson; $\bar u = 1-u$ is the momentum fraction carried by the
u or d antiquark. Note that one has to define $u$ in such a
way that the exponential  reads $\exp(i\ub\qz)$, rather than
$\exp(i u\qz)$, as it has to match the corresponding factor in
the hadronic parametrisation of the correlation function, cf.~Eq.~(\ref{eq:x}).
The $O(\alpha_s)$ corrections are new, and we find the following expressions:
\begin{eqnarray}
\pi_K & = & 1 + C_F\,\frac{\alpha_s}{4\pi}
      \left\{ \frac{51 - 2\,{\pi }^2}{3} +
        \ln \frac{-q^2}{\mu^2}
         \left( -\frac{9}{2}  - \ln\,u \right)  +
        \frac{2\,\left( -3 + 4\,u \right) \,\ln u}{\ub} -
        \frac{\ln^2 u}{\ub}\right.\nonumber\\
&&\left. +
        \left( -3 - 2\,\ln u \right) \,\ln \ub +
        2\,\ln^2\ub^2 + 4\,L_2(u) + 2\,L_2(-\frac{u}{\ub}) -
        2 \frac{u}{\ub}\left( L_2(\ub) - L_2(-\frac{\ub}{u})\right)
        \right\},
\nonumber\\
\pi_{\Kpar} & = & 1 + C_F\,\frac{\alpha_s}{4\pi}\left\{
      \frac{33-2\pi^2}{3} + 2 \ln^2\ub - (3+2\ln u)\ln \ub -
      \left(\frac{5}{2} + \ln u\right)
      \ln\,\frac{-q^2}{\mu^2}\right.\nonumber\\
&&\left. -
      \frac{2(3-4u)}{\ub}\,\ln u - \frac{\ln^2 u}{\ub} - \frac{2 u
      L_2(\ub)}{\ub} - \frac{2u L_2(-\ub/u)}{\ub} + 4 L_2(u) + 2
      L_2(-u/\ub)\right\}, \nonumber\\
\pi_{\Kperp} & = & 1 + C_F\,\frac{\alpha_s}{4\pi}\left\{ \frac{33
      -2\pi^2}{3} + 2 \ln^2\ub - 3 \,\frac{2-u}{\ub}\,\ln u -
      \frac{1+u}{\ub}\,\ln^2 u - \left( 3 + \frac{\ln u}{\ub}\right)
      \ln\,\frac{-q^2}{\mu^2}\right.\nonumber\\
&&\left. - \left( 3 + \frac{2 \ln u}{\ub} \right)
      \ln \ub - \frac{2u}{\ub}\, L_2(\ub) + 4 L_2(u) + 2 L_2 (-u/\ub) \right\}.
\end{eqnarray}
A collection of loop integrals necessary to perform these calculations
can be found in App.~\ref{app:c}.
We have checked that the above expressions reproduce the correct anomalous
dimensions for the Gegenbauer moments, $\gamma_{(n)}^\parallel$ and
$\gamma_{(n)}^\perp$, Eq.~(\ref{eq:1loopandim}). 
We have also checked that in the
local limit, i.e.\ $z\to 0$, $\exp(i\ub\qz)\to 1$, the expressions for
$\Kpar$ and $\Kperp$ agree: ${\cal
  P}\Pi_{\Kpar}\equiv {\cal P}\Pi_{\Kperp}$. Note that the sign of the
leading term differs with respect to \cite{Russians,CZreport}.

We have also calculated $O(\alpha_s)$ corrections to the condensate
terms and find
\begin{eqnarray}
{\cal P}\Pi_M^{\quark} & = & \frac{1}{q^2}\,\int_0^1 du\,e^{i\ub\qz} \left[
  \squark \left( \delta(u) + \frac{\alpha_s}{4\pi}\,C_F\left\{
  \delta(u)\left( 5\ln\,\frac{-q^2}{\mu^2} -
  3\right.\right.\right.\right.\nonumber\\
&& \left. + \left( 1 -
  \delta_{M,\parallel}\right) \left( 2 - 3 \delta_{M,\perp} + \left( 3
  \delta_{M,\perp} - 4 \right) \ln\,\frac{-q^2}{\mu^2}\right)\right) -
  2 \ub \left( 1 - \delta_{M,\perp}\right)\nonumber\\
&& \left.\left.\times \left( 1 + 2
  \delta_{M,\parallel} + \ln(u\ub) + \ln\,\frac{-q^2}{\mu^2}\right) +
  2 \left[ \frac{\ub}{u}\left( 2 - \ln(u\ub) - \ln\,\frac{-q^2}{\mu^2}
  \right) \right]_+\right\} \right)\nonumber\\
&& \left. + \quark ( u\leftrightarrow \ub)\vphantom{\frac{\alpha_s}{4\pi}}
  \right],
\label{eq:cond}
\end{eqnarray}
where the $[~~]_+$ prescription is defined as
$$[f(u)]_+ = f(u) - \delta(u_0) \int_0^1 dv\,f(v),$$
if $f$ has a simple pole (modulo logarithms) at $0\leq u_0\leq 1$.

For the contribution from the dimension 5 mixed condensates $\mixed$
and $\smixed$, we
restrict ourselves to the tree-level contribution and find
\begin{equation}\label{eq:cond2}
{\cal P}\Pi_M^{\mixed} = \frac{1}{3q^4}\int_0^1 du e^{i\ub \qz} \left[
  \smixed \{ (\delta_{M,\parallel} + \delta_{M,\perp}) \delta(u) +
  \delta'(u)\} + \mixed \{ u \leftrightarrow \ub\}\right].
\end{equation}

\section{Sum Rules for Moments}
\label{app:b}
\setcounter{equation}{0}

\begin{table}
\renewcommand{\arraystretch}{1.3}\addtolength{\arraycolsep}{4pt}
$$
\begin{array}{|l|lll|}
\hline
& \Gamma_1 & \Gamma_2 & \mbox{projection~}{\cal P}: \Pi_M\to {\cal
  P}\Pi_M\\\hline
K & \hat{z}\gamma_5 & \hat{z}\gamma_5 & 1/\qz^2\\
\Kpar & \hat{z} &  \hat{z} & 1/\qz^2\\[5pt]
\Kperp & \sigma_{\mu\nu}z^\nu & \sigma^{\mu\rho}z_\rho  &\displaystyle
\frac{1}{2-D}\,\frac{1}{\qz^2}\\[10pt]\hline
\end{array}
$$
\renewcommand{\arraystretch}{1}\addtolength{\arraycolsep}{-4pt}
\caption[]{Dirac structures and projections 
used for calculating even moments from
  diagonal correlation functions}\label{tab:B}
\end{table}

The Dirac structures and projections used for calculating the
diagonal correlation function Eq.~(\ref{eq:3.4}) are collected in
Tab.~\ref{tab:B}.

The nonperturbative corrections to the sum rule for even moments of
the $K$ can be extracted from \cite{CZreport}. We have
recalculated the SU(3) breaking terms explicitly and find agreement
with Ref.~\cite{CZreport}. The perturbative terms can be extracted
from Ref.~\cite{Gorsky}. 
We have also recalculated these terms in the nonlocal
operator formalism and confirm the result quoted in \cite{Gorsky}.
The complete sum rule for even Gegenbauer moments of the $K$ reads
\begin{eqnarray}
\lefteqn{\frac{3(n+1)(n+2)}{2(2n+3)}\,f_K^2\, a_n(\mu)
  e^{-m_K^2/M^2} = \frac{1}{2\pi^2}\,\frac{\alpha_s}{\pi}\,M^2
\left(1-e^{-s_0^\parallel/M^2}\right) }\nonumber\\
&&{}\times \int_0^1 du\,u\ub\,
C_n^{3/2}(2u-1)\,\ln^2\,\frac{u}{\ub}
+\frac{m_s\squark}{2M^2}\,(n+1)(n+2) \nonumber\\
&&{} + \frac{1}{24M^2}\,\gluon (n+1)(n+2) -
  \frac{m_s\smixed}{24M^4}\,n(n+1)(n+2)(n+3)\nonumber\\
&&{} +
  \frac{8\pi\alpha_s}{9}\,\frac{\quark\squark}{M^4}\,(n+1)(n+2) +
  \frac{4\pi\alpha_s}{81}\,\frac{\quark^2+\squark^2}{M^4}\,(n+1)^2(n+2)^2,
\end{eqnarray}
We have also checked the
nonperturbative terms in \cite{BBKT} and could not confirm the terms
in $m_s\smixed$. We thus find it appropriate to present here the
(hopefully) correct sum rules:
\begin{eqnarray}
\lefteqn{\frac{3(n+1)(n+2)}{2(2n+3)}\,(f^\parallel_{K})^2a_n^\parallel(\mu)
e^{-m_{K^*}^2/M^2}\ =}\nonumber\\
& = & \frac{1}{2\pi^2}\,\frac{\alpha_s}{\pi}\, M^2
\left(1-e^{-s_0^\parallel/M^2} \right) \int_0^1\!\! du\, u\bar u\,
C_n^{3/2}(2u-1)\,\,\ln^2\,\frac{u}{\ub}\nonumber\\
& & {}  +\frac{1}{2M^2}\,m_s\squark (n+1)(n+2) + \frac{1}{24M^2}\, \gluon
(n+1)(n+2) \nonumber\\
& & {}- \frac{1}{24M^4}\,m_s\smixed n(n+1)(n+2)(n+3)
-\frac{8\pi\alpha_s(\mu)}{9M^4}\,\quark\squark (n+1) (n+2)\nonumber\\
&& + \frac{4\pi\alpha_s}{81M^4}\,(\quark^2+\squark^2)(n+1)^2
(n+2)^2,
\end{eqnarray}
\begin{eqnarray}
\lefteqn{\frac{3(n+1)(n+2)}{2(2n+3)}\,(f_{K^*}^T(\mu))^2 a_n^\perp(\mu)
e^{-m_{K^*}^2/M^2}\ =}\nonumber\\
& = & \frac{1}{2\pi^2}\,\frac{\alpha_s}{\pi}\, M^2
\left(1-e^{-s_0^\perp/M^2} \right) \int_0^1\!\! du\, u\bar u\,
C_n^{3/2}(2u-1) \left( \ln u + \ln \bar u + \ln^2\,\frac{u}{\bar
u}\right)
\nonumber\\
& & {} + \frac{1}{24M^2}\, \gluon
(n^2+3n-2) - \frac{1}{24M^4}\,m_s\smixed (n+1)(n+2)(n^2+3n+4)\nonumber\\
& & {} +
\frac{4\pi\alpha_s}{81M^4}\,(\quark^2+\squark^2) (n-1)(n+1)(n+2)(n+4)+
\frac{1}{2M^2}\,m_s\squark (n+1)(n+2).\nonumber\\[-15pt]
\end{eqnarray}

Sum rules for odd
moments are obtained from the formulas given in App.~\ref{app:a} by
Borel transforming and replacing $\exp(i\bar u \qz)\to
C_n^{3/2}(2u-1)$. In this way, and using the properties of
Gegenbauer polynomials as for instance collected in \cite{PT}, which
amount to the following replacements:
$$
\begin{array}{l@{\quad}l}
\displaystyle\delta(u)\to (-1)^n\, \frac{1}{2}\,(n+1)(n+2), & 
\displaystyle\delta(\ub)\to \frac{1}{2}\,(n+1)(n+2),\\[10pt]
\displaystyle\delta'(u)\to (-1)^n\, \frac{1}{4}\,n(n+1)(n+2)(n+3), & 
\displaystyle\delta'(\ub)\to \frac{1}{4}\,n(n+1)(n+2)(n+3),\\
\end{array}
$$
we
find for the $O(\alpha_s^0)$ contributions for odd $n$:
\begin{eqnarray}
{\cal B}_{\mbox{\scriptsize sub}}{\cal P}\Pi^{(n)}_{K} & = &
-\frac{3}{8\pi^2}\,m_s\, \left( 1 - e^{-s_0/M^2}\right) +
\frac{1}{2M^2}\,(n+1) (n+2) (\squark - \quark)\nonumber\\
&& {}+
\frac{1}{12M^4}\,n(n+1) (n+2)(n+3) (\mixed - \smixed)\,,\nonumber\\
{\cal B}_{\mbox{\scriptsize sub}}{\cal P}\Pi^{(n)}_{\Kpar} & = &
-\frac{3}{8\pi^2}\,m_s\, \left( 1 - e^{-s_0/M^2}\right) +
\frac{1}{2M^2}\,(n+1) (n+2) (\squark - \quark)\nonumber\\
&&{} +
\frac{1}{12M^4}\,(n+1)^2 (n+2)^2 (\mixed - \smixed)\,,\nonumber\\
{\cal B}_{\mbox{\scriptsize sub}}{\cal P}\Pi^{(n)}_{\Kperp} & = &
{\cal B}_{\mbox{\scriptsize sub}}{\cal P}\Pi^{(n)}_{\Kpar}\,.
\end{eqnarray}
Note that the sign for the leading order perturbative
contribution is different with respect to (C.9) and (C.10) in
Ref.~\cite{BBKT} and also with respect to \cite{Russians,CZreport}. We
  have carefully checked that we obtain the correct sign in the local
  limit $z\to 0$ and that our result stays unchanged for a different
  choice of the position of the nonlocal current in coordinate space, for
  instance $\bar s(0)\Gamma_1[0,z]q(z)\to \bar s(-z)\Gamma_1[-z,z]q(z)$.
As it turns out, also the mixed condensate-contributions differ from
the formulas given in \cite{BBKT}.

To obtain the $O(\alpha_s)$ corrected sum rules for odd moments from
the formulas given in App.~\ref{app:a}, we also need the following
continuum-subtracted Borel transforms:
\begin{eqnarray*}
{\cal B}_{\mbox{\scriptsize
      sub}}\,\ln\,\frac{-q^2-i0}{\mu^2} &=& 
-\left(1-e^{-s_0/M^2}\right),\\
 {\cal B}_{\mbox{\scriptsize
      sub}}\,\ln^2\,\frac{-q^2-i0}{\mu^2} &=&
      -\,\frac{1}{M^2}\int_0^{s_0}ds\,e^{-s/M^2}
      \,2\,\ln\,\frac{s}{\mu^2}\,,\\
{\cal B}_{\mbox{\scriptsize
      sub}}\,\frac{1}{q^2}\,\ln\,\frac{-q^2-i0}{\mu^2} & = &
      \frac{1}{M^2}\left\{ \gamma_E - \ln\,\frac{M^2}{\mu^2} - {\rm
      Ei}\left(-\,\frac{s_0}{M^2}\right)\right\},
\end{eqnarray*}
which complete the set of formulas needed to translate the
correlation functions obtained in the previous appendix into QCD sum
rules for odd moments.

\section{Loop Integrals}
\label{app:c}
\setcounter{equation}{0}

\begin{figure}
$$\epsfxsize=0.45\textwidth\epsffile{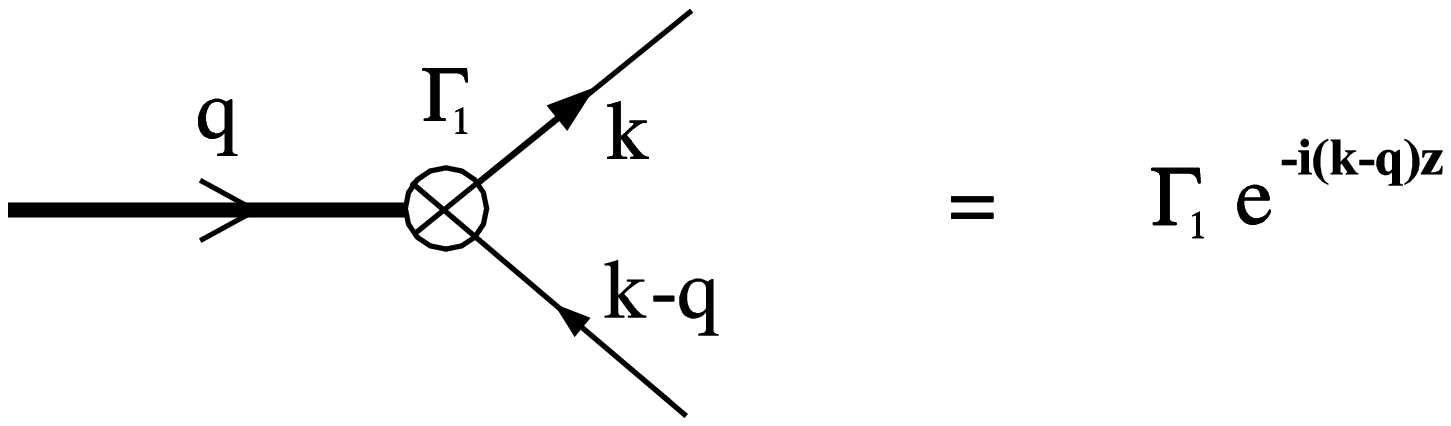}\quad
\epsfxsize=0.45\textwidth\epsffile{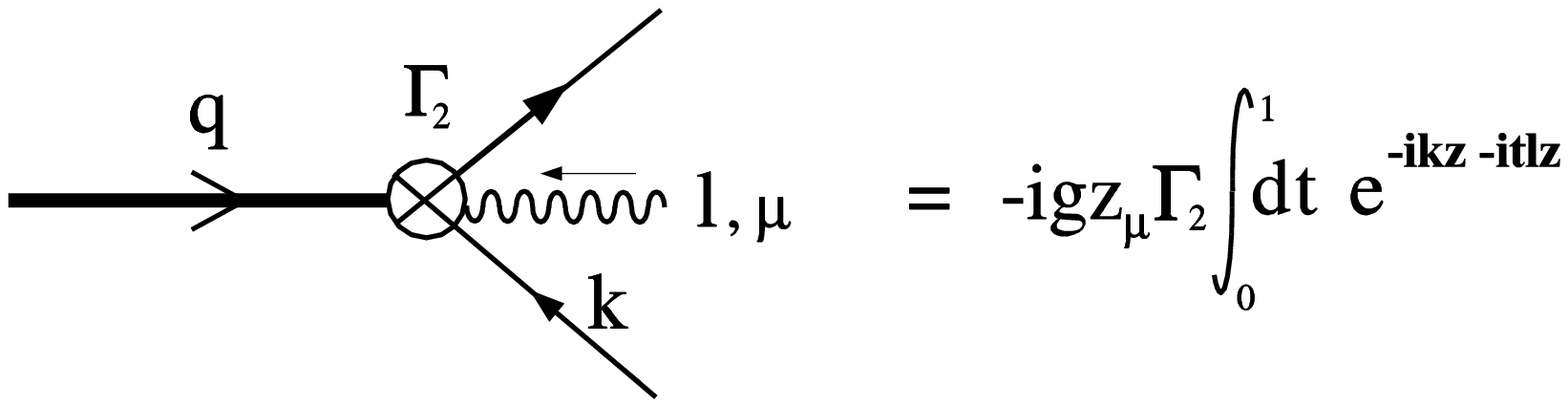}$$ 
\caption[]{$O(g_s^0)$ and $O(g^1_s)$ Feynman rules induced by
  the nonlocal current in (\ref{eq:CF}).}\label{fig:A}
\end{figure}

To the benefit of apprentices, and also for future reference, we
collect in this appendix relevant one- and two-loop integrals. The
main difference as compared to usual calculations is the
nonlocal vertex induced by $\bar s(0)\Gamma_1 [0,z] q(z)$, which, to
order $\alpha_s$, gives rise to the Feynman rules shown in
Fig.~\ref{fig:A}.

The master one-loop integral is given by ($z^2=0$, $D=4+2\epsilon$):
\begin{eqnarray}
\int \left[ d^Lk\right] \,e^{-i(k-q)z} \,\frac{(kz)^a}{(k^2)^\alpha
  ((k-q)^2)^\beta} & = & (-1)^{\alpha+\beta}
  (-q^2)^{2-\alpha-\beta+\epsilon} \qz^a
  \,\frac{\Gamma(\alpha+\beta-2-\epsilon)}{\Gamma(\alpha)\Gamma(\beta)}
\nonumber\\ 
  &&\times\int_0^1 e^{iu\qz} \,u^{1-\beta+\epsilon} \ub^{1-\alpha+a+\epsilon},
\end{eqnarray}
where $[d^Lk]\, i/(4\pi)^{D/2}  \equiv d^Dk/(2\pi)^D$.

As for two-loop integrals, one needs for instance
\begin{eqnarray}
\lefteqn{
\int_0^1\!\! dv\! \int \!\left[d^Lk\right] \left[d^Ll\right] e^{-i(k-qz)}\,
e^{-iv(l-k)z}\,\frac{(kz)^a(lz)^b}{k^2l^2(k-l)^2(k-q)^2} =
\frac{\Gamma(-\epsilon)\Gamma(-2\epsilon)}{\Gamma(1-\epsilon)}\,\qz^{a+b} 
(-q^2)^{2\epsilon}}\hskip3.5cm \nonumber\\
&&\times \frac{1}{i\qz}\, \int_0^1 du\,
e^{iu\qz} \left\{ u^{2\epsilon} \ub^{b+\epsilon} \int_0^1
dy\,y^\epsilon \bar y^{-1+\epsilon} \left( (1-u y)^{a-1} -
\ub^{a-1}\right)\right.\nonumber\\
&&\left. +
\frac{\Gamma(\epsilon)\Gamma(1+\epsilon)}{\Gamma(1+2\epsilon)}\,
u^{2\epsilon} \ub^{a+b+\epsilon-1} - \frac{\Gamma(\epsilon)\Gamma(b+1+
\epsilon)}{\Gamma(b+1+2\epsilon)}\,u^\epsilon\ub^{a+b-1+2\epsilon} \right\}.
\end{eqnarray}
Generally, all loop-integrals with additional exponentials can be
calculated conveniently using Feynman parameters. The calculation is
further simplified by the fact that one only needs the imaginary part
 in $q^2$, which implies that finite integrals need not be
calculated. ``Overlapping exponentials'' like for instance
$$\int_0^1 dx \int_0^1 dy\, \exp(i(x+y\bar x)\qz) \,f(x,y)$$
can be rewritten as
$$\int_0^1 du\,\exp(i\ub \qz) \int_0^1 dx \int_0^1 dy\, \delta(x+y\bar
x-\ub) f(x,y),$$
which allows one to reduce all contributions to the canonical form.

\end{document}